\documentclass[prd,showpacs,floatfix,twocolumn,amsmath,amssymb,floatfix,nofootinbib]{revtex4}
\usepackage{graphicx,color,dcolumn,booktabs,bm}
\usepackage{longtable,lscape}
\usepackage{txfonts}
\usepackage{overpic}
\usepackage{epsfig}
\usepackage{amssymb}
\usepackage{rotating}
\usepackage{epstopdf}
\usepackage{indentfirst}
\usepackage{feynmf}   
\usepackage{slashed}  
\usepackage{cases}
\usepackage{color}
\usepackage{multirow}
\usepackage{graphicx,color,dcolumn,booktabs,bm}

\graphicspath{{Figures/}} %

\graphicspath{{Figures/}} %
\usepackage[colorlinks, citecolor=blue,anchorcolor=red,menucolor=red, linkcolor=red,filecolor=red,runcolor=red,frenchlinks=red]{hyperref}
\usepackage{cases}

\begin{document}

\title{Approximate degeneracy of heavy-light mesons with the same $L$}
\author{Takayuki Matsuki$^{1,2}$}
\email{matsuki@tokyo - kasei.ac.jp}
\author{Qi-Fang L\"u$^{3}$}\email{lvqifang@ihep.ac.cn}
\author{Yubing Dong$^{3,4}$}\email{dongyb@ihep.ac.cn}
\author{Toshiyuki Morii$^{5}$}\email{morii@kobe-u.ac.jp}
\affiliation{$^1$Tokyo Kasei University, 1-18-1 Kaga, Itabashi, Tokyo 173-8602, Japan\\
$^2$Theoretical Research Division, Nishina Center, RIKEN, Wako, Saitama 351-0198, Japan\\
$^3$Institute of High Energy Physics, CAS, Beijing 100049, People's Republic of China\\
$^4$Theoretical Physics Center for Science Facilities (TPCSF), CAS, People's Republic of China\\
$^5$Faculty of Human Development, Kobe University, 3-11 Tsurukabuto, Nada, Kobe 657-8501, Japan}

\begin{abstract}

Careful observation of the experimental spectra of heavy-light mesons tells us that heavy-light mesons with the same angular momentum $L$ are almost degenerate. The estimate is given how much this degeneracy is broken in our relativistic potential model, and it is analytically shown that expectation values of a commutator between the lowest order Hamiltonian and ${\vec L}^{~2}$ are of the order of $1/m_Q$ with a heavy quark mass $m_Q$. It turns out that nonrelativistic approximation of heavy quark system has a rotational symmetry and hence degeneracy among states with the same $L$. This feature can be tested by measuring higher orbitally and radially excited heavy-light meson spectra for $D/D_s/B/B_s$ in LHCb and forthcoming BelleII.

\end{abstract}

\maketitle

\section{introduction}\label{sec1}

Ever since the discovery of $X(3872)$, $D_{s0}(2317)$, and $D_{s1}(2460)$ in 2003, there have been many more $XYZ$ as well as higher radially and orbitally excited particles found at Belle, BESII, BESIII, BaBar, and LHCb \cite{Beringer:1900zz}. There are a couple of problems for these particles. One is that most of them appear at thresholds and hence there may be kinematical explanations possible. Another point is that some of them should
be multiquark states because they cannot be explained as higher excited states of ordinary quarkonium due to the charged states.

When focusing on higher orbital excitations of the heavy-light system, we see some tendency of their spectroscopy which has not yet been explained by heavy quark symmetry. The problem is described as follows. Even though the angular momentum $L$ is not a good quantum number in the heavy quark system, it seems that masses of states with the same $L$ are close to each other even for the heavy-light system.

To explain this approximate degeneracy among heavy-light mesons with the same $L$ observed in experiments, we need to show, at least analytically or numerically, how small matrix elements of this resultant difference operator are. One of the powerful quark models is the relativized Godfrey-Isgur (GI) model \cite{Godfrey:1985xj, Godfrey:1986wj} in which their lowest order Hamiltonian commutes with $\vec L$ even in their relativized formulation. Hence, there is no wonder within their formulation why the masses with the same $L$ are close to each other.
However, when calculating commutator of the lowest order Hamiltonian and $\vec L$ in our relativistic potential model \cite{Matsuki:1997da,mms}, we obtain nonvanishing result.  Difference between the GI and our models is in that we cast a light quark into a four-component Dirac spinor which causes non-vanishing commutator as seen below while the GI treats it a two-component spinor.

In the past decades, the heavy-light meson families have become a rich structure as seen in PDG \cite{Beringer:1900zz}. Even though it does not take into account the heavy quark symmetry, the GI model \cite{Godfrey:1985xj, Godfrey:1986wj} has been successful in reproducing and predicting low lying hadrons and heavy-light mesons except for $D_{sJ}$. This model respects angular momentum conservation at the lowest order so that states with the same angular momentum $L$ are degenerate without spin-orbit interactions.

Let us look at numerical results of models only for $D$ mesons which include a heavy quark $c$ and compare them with each other and with experimental data in Table \ref{table1}. A model in the {second column \cite{Godfrey:1985xj,gi2,gi3} is the GI model itself and a model in the seventh column \cite{ls1,ls2} is a nonrelativistic potential model including a one-loop computation of the heavy-quark interaction}. Those in the {third} column \cite{zvr,de} use the Bethe-Salpeter formulation to expand the system in terms of $1/m_Q$, while ours in the {sixth} column \cite{mms} uses the Foldy-Wouthouysen-Tani transformation to obtain the equation of motion for a $Q\bar q$ bound system and is essentially the same formulation as that of Ref. \cite{zvr}. Hence the following arguments given in Sect. \ref{sec2} can be derived from Refs. \cite{zvr,de}, too. Finally Ref. \cite{efg} uses a quasipotential approach whose details are given in their paper.
Similar tables for $D_s/B/B_s$ mesons can be easily obtained and they give tendency similar to Table \ref{table1}. Because we would like to extract and show the essence of our claim, we omit them in this article. It is not amazing to see that states with the same $L$ of the GI model have similar mass values for states with the same $L$ because it respects $L$. However, it is surprising that even models respecting heavy quark symmetry produce the results similar to the GI model, which can be seen from Table~\ref{table1}.

States in Table \ref{table1} are assigned definite values of $^{2S+1}L_J$ {in the first column}. Even though our relativistic wave function is not an egenstate of $L$ in our formulation \cite{mms}, we can still assign $^{2S+1}L_J$ to each state in the nonrelativistic limit.

In the last two columns of Table \ref{table1}, we give average values of experimental data within a spin doublet of the heavy-quark system and gap values between spin doublets. 
{For instance, average values are given by 1938 MeV for a spin multiplet $(J^P=0^-,1^-)$, 2394 MeV for a multiplet $(0^+,1^+)$, 2443 MeV for a multiplet $(1^+,2^+)$, 2763 MeV for a multiplet $(1^-,2^-)$, etc. Gap values are given by difference of these values, i.e., 456 MeV between multiples $(0^-,1^-)$ with $L=0$ and $(0^+,1^+)$ with $L=1$, 49 MeV between $(0^+,1^+)$ and $(1^+,2^+)$ with the same $L=1$, 320 MeV between $(1^+,2^+)$ with $L=1$ and $(1^-,2^-)$ with $L=2$, etc.} 
We can see that mass differences within a spin doublet and {between} doublets with the same $L$ are very small compared with a mass gap between different multiplets with different $L$, which is nearly equal to the value of the QCD $\Lambda_{QCD}\sim 300$ MeV{\footnote{We expect that a gap value is roughly $\Lambda_{QCD}\sim 300$ MeV because this gap is caused by strong interaction characterized by $\Lambda_{QCD}$, which is numerically shown in Ref. \cite{Matsuki:2007iu} when deriving mass gap relation between two spin multiplets. In Ref. \cite{Beringer:1900zz}, the notation $\Lambda_{\overline{MS}}$ is taken instead of $\Lambda_{QCD}$.} \cite{Beringer:1900zz} for $n_f=4$.

\newcommand{\Slash}[1]{\ooalign{\hfil/\hfil\crcr$#1$}}
\begin{table*}[htbp]
\begin{center}
\caption{ \label{tmass1} The $D$ meson masses in MeV from different quark models and experimental data. Models of ZVR\cite{zvr}, DE\cite{de}, EFG\cite{efg}, and MMS\cite{mms}
respect heavy-quark symmetry. \label{table1}}
\footnotesize
\begin{tabular}{lcccccccccc}
\hline\hline
  State            & GI\cite{Godfrey:1985xj,gi2,gi3}  & ZVR\cite{zvr}    & DE\cite{de}     &EFG\cite{efg}  &MMS\cite{mms} &LS\cite{ls1,ls2}  & EXP\cite{exp1,exp2,exp3,exp4,exp6}    & Average            & Gap   \\\hline
  $D(1^1S_0)$      & 1874       & 1850          & 1868            & 1871      & 1869        &1867                & 1867  & \multirow{2}{*}{1938}     &        \\
  $D(1^3S_1)$      & 2038       & 2020          & 2005            & 2010      & 2011        &2010                & 2009   &                           &         \\\hline
  $D(1^3P_0)$      & 2398       & 2270          & 2377            & 2406      & 2283        &2252                & 2361  & \multirow{2}{*}{2394}     & \multirow{2}{*}{456}  \\
  $D_1(1P)$        & 2455       & 2400          & 2417            & 2426      & 2421        &2402                & 2427 &                           &           \\
  $D^\prime_1(1P)$ & 2467       & 2410          & 2490            & 2469      & 2425        &2417                & 2422  & \multirow{2}{*}{2443}     & \multirow{2}{*}{49}  \\
  $D(1^3P_2)$      & 2501       & 2460          & 2460            & 2460      & 2468        &2466                & 2463 &                           &              \\\hline
  $D(1^3D_1)$      & 2816       & 2710          & 2795            & 2788      & 2762        &2740                & 2781 & \multirow{2}{*}{2763}     & \multirow{2}{*}{320} \\
  $D_2(1D)$        & 2816       & 2740          & 2775            & 2806      & 2800        &2693                & 2745 &                           &        \\
  $D^\prime_2(1D)$ & 2845       & 2760          & 2833            & 2850      & $-$         &2789                & 2745  & \multirow{2}{*}{2763}     & \multirow{2}{*}{0}      \\
  $D(1^3D_3)$      & 2833       & 2780          & 2799            & 2863      & $-$         &2719                & 2800/2762   &             &             \\\hline
  $D(1^3F_2)$      & 3132       & 3000          & 3101            & 3090      & $-$         &$-$                 & $-$ &             &         \\
  $D_3(1F)$        & 3109       & 3010          & 3074            & 3129      & $-$         &$-$                 & $-$ &             &          \\
  $D^\prime_3(1F)$ & 3144       & 3030          & 3123            & 3145      & $-$         &$-$                 & $-$ &             &         \\
  $D(1^3F_4)$      & 3113       & 3030          & 3091            & 3187      & $-$         &$-$                 & $-$ &             &          \\\hline
  $D(1^3G_3)$      & 3398       & 3240          &$-$              & 3352      & $-$         &$-$                 & $-$ &             &         \\
  $D_4(1G)$        & 3365       & 3240          &$-$              & 3403      & $-$         &$-$                 & $-$ &             &         \\
  $D^\prime_4(1G)$ & 3400       & 3260          &$-$              & 3415      & $-$         &$-$                 & $-$ &             &         \\
  $D(1^3G_5)$      & 3362       & $-$           &$-$              & 3473      & $-$         &$-$                 & $-$ &             &         \\\hline
  $D(2^1S_0)$      & 2583       & 2500          & 2589            & 2581      & $-$         &2555                & 2560    &\multirow{2}{*}{2595}     &                    \\
  $D(2^3S_1)$      & 2645       & 2620          & 2692            & 2632      & $-$         &2636                & 2692&             &                    \\\hline
  $D(2^3P_0)$      & 2932       & 2780          & 2949            & 2919      & $-$         &2752                & $-$ &             &             \\
  $D_1(2P)$        & 2925       & 2890          & 2995            & 2932      & $-$         &2886                & $-$ &             &             \\
  $D^\prime_1(2P)$ & 2961       & 2890          & 3045            & 3021      & $-$         &2926                & $-$  &             &            \\
  $D(2^3P_2)$      & 2957       & 2940          & 3035            & 3012      & $-$         &2971                & $-$ &             &                    \\\hline
  $D(2^3D_1)$      & 3232       & 3130          & $-$             & 3228      & $-$         &3168                & $-$ &             &             \\
  $D_2(2D)$        & 3212       & 3160          & $-$             & 3259      & $-$         &3145                & $-$ &             &              \\
  $D^\prime_2(2D)$ & 3249       & 3170          & $-$             & 3307      & $-$         &3215                & $-$ &             &             \\
  $D(2^3D_3)$      & 3227       & 3190          & $-$             & 3335      & $-$         &3170                & $-$ &             &          \\\hline
  $D(2^3F_2)$      & 3491       & 3380          & $-$             &$-$        & $-$         &$-$                 & $-$&             &             \\
  $D_3(2F)$        & 3462       & 3390          & $-$             &$-$        & $-$         &$-$                 & $-$ &             &           \\
  $D^\prime_3(2F)$ & 3499       & 3410          & $-$             &$-$        & $-$         &$-$                 & $-$&             &            \\
  $D(2^3F_4)$      & 3466       & 3410          & $-$             & 3610      & $-$         &$-$                 & $-$&             &            \\\hline
  $D(2^3G_3)$      & 3722       & $-$           & $-$             & $-$       & $-$         &$-$                 & $-$ &             &           \\
  $D_4(2G)$        & 3687       & $-$           & $-$             & $-$       & $-$         &$-$                 & $-$&             &            \\
  $D^\prime_4(2G)$ & 3723       & $-$           & $-$             & $-$       & $-$         &$-$                 & $-$ &             &           \\
  $D(2^3G_5)$      & 3685       & $-$           & $-$             & 3860      & $-$         &$-$                 & $-$ &             &          \\
  \hline\hline

\end{tabular}
\end{center}
\end{table*}

\section{Analytical analysis}\label{sec2}

Using the heavy quark symmetry, the lowest order Hamiltonian in our relativistic potential model \cite{Matsuki:1997da,mms} is given by
\begin{eqnarray}
  H_0=\vec\alpha_q\cdot\vec p + m_q\beta_q,
\end{eqnarray}
whose commutation relation with $\vec L=\vec r\times \vec p$ is given by
\begin{eqnarray}
  \left[H_0, L_i\right] = -i\left(\vec\alpha_q\times\vec p\right)_i. \label{eq:moment}
\end{eqnarray}
On the other hand, we have the following commutation relation,
\begin{eqnarray}
  \left[H_0,\frac{1}{2}\Sigma_{qi}\right] = i\left(\vec\alpha_q\times\vec p\right)_i, \label{eq:spin}
\end{eqnarray}
with a light quark spin $\vec\Sigma_q/2$. Adding Eqs. (\ref{eq:moment}) and (\ref{eq:spin}), we obtain conservation of $\vec j_\ell=\vec L+\vec\Sigma_q/2$ of light-quark degrees of freedom as expected, $\left[H_0,\vec j_\ell\right] = 0$.
Because matrices related to a heavy quark are not included in $H_0$,  a heavy quark spin $\vec\Sigma_Q/2$ also commutes with $H_0$, $\left[H_0,\vec\Sigma_Q/2\right]=0$, which means a total angular momentum $\vec J = \vec L+\vec\Sigma_q/2+\vec\Sigma_Q/2$ also conserves,  $\left[H_0,\vec J\right] = 0$.

We would like to estimate the expectation value of $[H_0,\vec L^2]$
whose explicit form is given by
\begin{eqnarray}
  \mathcal{M}=[H_0,\vec L^2]=i\alpha_{qj}\left(ip_j - r_j p^2 + (r\cdot p)p_j \right) \equiv \alpha_{qj}f_j(r,p).
  \label{eq:M}
\end{eqnarray}

There is a lemma that if we calculate the expectation value, $\int \Psi_\ell^\dag\left[H_0,\mathcal{O}\right]\Psi_\ell$, and if $\Psi_\ell$ is an eigenfunction of $H_0$ with a real eigenvalue $E_\ell$, i.e., $H_0\Psi_\ell=E_\ell\Psi_\ell$, then $\int \Psi_\ell^\dag\left[H_0,\mathcal{O}\right]\Psi_\ell=0$ because $\int \Psi_\ell^\dag\left[H_0,\mathcal{O}\right]\Psi_\ell=\int \Psi_\ell^\dag \left(E_\ell \mathcal{O} - \mathcal{O} E_\ell \right)\Psi_\ell=0$ for any operator $\mathcal{O}$.

The actual wave function includes both positive- and negative-energy states, $\Psi_{\ell}^\pm$ in regard to a heavy quark,
\begin{eqnarray}
  \psi_\ell = \Psi_\ell^+ + \sum_{\ell'} \left(c^{\ell,\ell'}_+\Psi_{\ell'}^+ +
  c^{\ell,\ell'}_-\Psi_{\ell'}^-\right), \label{eq:psi_ell}
\end{eqnarray}
where $\ell=\left\{k,j,m\right\}$ with a total angular momentum $j$ and its $z$-component $m$. Here the quantum number $k$ is related to the angular momentum of a light quark $j_\ell$ and the parity $P$ for a heavy-light meson as \cite{Matsuki:1997da,Matsuki:2004zu},
\begin{eqnarray}
  j_\ell &=& |k|-\frac{1}{2},\quad P=\frac{k}{|k|}(-1)^{|k|+1}.
\end{eqnarray}
Wave functions are defined as \cite{Matsuki:1997da},
\begin{eqnarray}
  &&\Psi_{\ell}^{+} = \left(0\; \psi_{jm}^k\right),\quad
  \Psi_{\ell}^{-}=\left(\psi_{jm}^k\; 0\right),\nonumber \\
  &&\psi_{jm}^k(r,\Omega) = \frac{1}{r}
  \left( {\begin{array}{*{20}{c}}
  {u_k(r)y_{jm}^k}\\ {iv_k(r)y_{jm}^{ - k}} \end{array}} \right). \label{eq:psi}
\end{eqnarray}
In the case of $k=-1$ ($j_\ell^P=(1/2)^-$), we obtain the following results up to the first order of $1/m_Q$, \cite{Matsuki:1997da,Matsuki:2007fb},
\begin{eqnarray}
  \psi_\ell(0^-) &=& \Psi_{-1}^+ + c_{1-}^{-1,1}\Psi_1^- + O\left(1/m_Q^2\right), \\
  \psi_\ell(1^-) &=& \Psi_{-1}^+ + c_{1+}^{-1,2}\Psi_2^+ + c_{1-}^{-1,-2}\Psi_{-2}^- + O\left(1/m_Q^2\right),
\end{eqnarray}
where we give $J^P$ in the parentheses on the l.h.s. and all the constants, $c_{1\pm}^{k,k'}$, are of the order of $1/m_Q$. On the r.h.s there appear a wave  function with a negative-energy component of a heavy quark, $\Psi^-$, together with a positive energy one, $\Psi^+$.
After some calculations, we obtain the matrix elements,
\begin{eqnarray}
  &&\left<\Psi^+_{\ell'}|\mathcal{M}|\Psi^-_\ell\right> =
  \left<\Psi^-_{\ell'}|\mathcal{M}|\Psi^+_\ell\right> =0, \label{com+-}
  \\
  &&\left<\Psi^\pm_{\ell'}|\mathcal{M}|\Psi^\pm_\ell\right> =
  i\int d^3r
  \frac{1}{r}\;\Biggl[-{v_{k'}}(r)y_{j'm'}^{ - k'\dag }\sigma_n f_n(r,p)
  \nonumber \\
  &&\left.\times\left(\frac{1}{r}
  {u_k}(r)y_{jm}^k \sigma_i\right)
  +{u_{k'}}(r)y_{j'm'}^{k'\dag }\sigma_n f_n(r,p)\left(\frac{1}{r}
  {v_k}(r)y_{jm}^{ -k } \sigma_i\right) \right],
  \nonumber \\
  \label{com++--}
\end{eqnarray}
where $\sigma_i$'s are Pauli matrices, $p_i$ is a momentum operator, and $f_n(r,p)$ is defined in Eq. (\ref{eq:M}).

When estimating $\left<\psi_\ell(0^-)|\mathcal{M}|\psi_\ell(0^-)\right>$, there is no surviving term up to the first order in $1/m_Q$. This is because $\left<\Psi^+_{-1}|\mathcal{M}|\Psi^+_{-1}\right>$ vanishes due to the lemma even though we have Eq. (\ref{com++--}) and cross terms of $\Psi_\ell^+$ and $\Psi_{\ell'}^-$ vanish because of Eq. (\ref{com+-}). Hence, the surviving term starts from the order of $(1/m_Q)^2$. When estimating $\left<\psi_\ell(1^-)|\mathcal{M}|\psi_\ell(1^-)\right>$ and taking into account the above estimate and $c_{1-}^{-1,1}\sim 1/m_Q$, there remain cross terms in $k$, $\left<\Psi^+_{-1}|\mathcal{M}|\Psi^+_{2}\right>$ with $k$ quantum numbers in subindices and its conjugate, which are of the order of $1/m_Q$ and hence it is suppressed for large $m_Q$. The similar arguments for other higher states give the same conclusion and the expectation value of a matrix element for a higher state is all the same order of magnitude, i.e., at most $1/m_Q$.

In order to obtain a complete symmetry, we just need to neglect a lower component radial wave function $v_k(r)$ which makes Eq.~(\ref{com++--}) vanish. Neglecting $v_k(r)$ in Eq.~(\ref{eq:psi}), we obtain a nonrelativistic wave function in the heavy quark system and a little calculation shows us that this is an eigenfunction of $\vec L~^2$ as,
\begin{eqnarray}
  \vec L~^2 y_{jm}^k = k(k+1) y_{jm}^k = L(L+1) y_{jm}^k, \label{eq:LL}
\end{eqnarray}
where use has been made of a formula, $\vec L\cdot \sigma_q\otimes y_{jm}^k = -(k+1) y_{jm}^k$ and the fact that $k=L$ or $-(L+1)$. Inclusion of a radial wave function does not change the result because $\vec L\; u_k(r)=0$.
Eq. (\ref{eq:LL}) means that nonrelativistic approximation of the heavy quark system has a rotational symmetry and hence in this approximation states with the same $L$ are degenerate.

\section{Conclusions and discussion}\label{sec4}

In this article, we have pointed out that there exists an approximate degeneracy among heavy-light systems with the same $L$.
This is supported by an experimental fact which can be seen from Table \ref{table1}. This approximate symmetry explains why the GI model obtains results similar to those of the heavy-light systems which are fitted well with experimental data. This is because the GI model has this symmetry from the beginning which is broken by the spin-orbit interactions. Numerical results of the GI model together with those of other models respecting heavy quark symmetry have been compared with the experimental data of the $D$ mesons in Table~\ref{table1} and they well give similar results to each other.

We have analytically shown that expectation values of $[H_0,\vec L^{~2}]$ give us at most of the order of $1/m_Q$ for $0^-$ and $1^-$ states and the similar arguments will give us the same conclustion for other higher states in our model which respects heavy quark symmetry. Note that this order of magnitude, $1/m_Q$, is the same as those which break degeneracy of a spin doublet of heavy-light systems. It has been shown that there is a rotational symmetry in the limit of $m_Q\to \infty$ and nonrelativistic limit of heavy-quark symmetry as shown in Eq.~(\ref{eq:LL}).

Simple application of our idea to other states can be given by baryons $QQq$ like $\Xi_{cc}^+$, multiquark states in which one light quark is included like $QQ\bar Qq$, and probably other states in which a couple of light quarks can be regarded as a brown mock. {A good expample is given by a spectrum of $\Lambda_c$ which gives us $\Lambda_c(2286)$ with $L=0$, $\Lambda_c^+(2595)$ and $\Lambda_c^+(2625)$ with $L=1$, and $\Lambda_c^+(2880)$ and $\Lambda_c^+(2940)$ with $L=2$ \cite{Beringer:1900zz}, where a spin multiplet is given by member/members with the same $L$. $L$ is defined by an angular momentum between a heavy quark $c$ and two light quarks $(ud)$. One can easily see that gaps between different spin multiplets are nearly equal to $\Lambda_{QCD}\sim 300$ MeV, which coincides with the observation of heavy-light mesons.}

Future measurement of higher orbitally and/or radially excited states and their masses by LHCb and forthcoming BelleII is waited for to test our observation.\\


\section*{Acknowledgements}

This work is partly supported by the National
Natural Science Foundation of China under Grant
No. 11475192, as well as supported, in part, by the
DFG and the NSFC through funds provided to the Sino-German
CRC 110 Symmetries and the Emergence of Structure in QCD.
T. Matsuki wishes to thank Yubing Dong for his kind hospitality at IHEP, Beijing,
where part of this work was carried out, and also K. Seo for helpful discussions.


\begin{thebibliography}{50}
\bibitem{Beringer:1900zz}
 K.~A.~Olive {\it et al.}  [Particle Data Group Collaboration],
  Chin.\ Revs.\ C {\bf 38}, 090001 (2014).

\bibitem{Godfrey:1985xj}
  S.~Godfrey and N.~Isgur,
  Phys.\ Rev.\ D {\bf 32} (1985) 189.

\bibitem{Godfrey:1986wj}
  S.~Godfrey and R.~Kokoski,
  Phys.\ Rev.\ D {\bf 43} (1991) 1679.

\bibitem{mms}
  T.~Matsuki, T.~Morii and K.~Sudoh,
  Prog.\ Theor.\ Phys.\  {\bf 117}, 1077 (2007).

\bibitem{Matsuki:1997da}
  T.~Matsuki and T.~Morii,
  Phys.\ Rev.\ D {\bf 56}, 5646 (1997)
  [hep-ph/9702366].


\bibitem{gi2}
  Q.~F.~L\"{u} and D.~M.~Li,
  Phys.\ Rev.\ D {\bf 90}, no. 5, 054024 (2014).

\bibitem{gi3}
  Y.~Sun, X.~Liu and T.~Matsuki, Phys.\ Rev.\ D {\bf 88}, 094020 (2013).
\bibitem{ls1}
  O.~Lakhina and E.~S.~Swanson,
  Phys.\ Lett.\ B {\bf 650}, 159 (2007).

\bibitem{ls2}
  D.~-M.~Li, P.~-F.~Ji and B.~Ma,
  Eur.\ Phys.\ J.\ C {\bf 71}, 1582 (2011).
\bibitem{ls1}
  O.~Lakhina and E.~S.~Swanson,
  Phys.\ Lett.\ B {\bf 650}, 159 (2007).

\bibitem{ls2}
  D.~-M.~Li, P.~-F.~Ji and B.~Ma,
  Eur.\ Phys.\ J.\ C {\bf 71}, 1582 (2011).

\bibitem{zvr} J. Zeng, J. W. Van Orden, and W. Roberts, Phys. Rev. D {\bf 52}, 5229 (1995).

\bibitem{de} M. Di Pierro and E. Eichten, Phys. Rev. D  {\bf 64}, 114004 (2001).

\bibitem{efg} D. Ebert, R. N. Faustov, and V. O. Galkin, Eur. Phys. J. C {\bf 66}, 197 (2010).

\bibitem{exp1}
  K.~A.~Olive {\it et al.}  [Particle Data Group Collaboration],
  Chin.\ Phys.\ C {\bf 38}, 090001 (2014).

\bibitem{exp2} R. Aaij et al. (LHCb Collaboration), J. High Energy Phys. {\bf 1309}, 145 (2013).

\bibitem{exp3}  R. Aaij et al. (LHCb Collaboration), Phys. Rev. D {\bf 91}, 092002 (2015).

\bibitem{exp4}        R. Aaij et al. (LHCb Collaboration), Phys. Rev. D {\bf 92}, 032002 (2015).

\bibitem{exp6}        P. del. Amo Sanchez et al. (BaBar Collaboration), Phys. Rev. D {\bf 82}, 111101 (2010).
\bibitem{Matsuki:2007iu} 
  T.~Matsuki, T.~Morii and K.~Sudoh,
  Phys.\ Lett.\ B {\bf 659}, 593 (2008)
  doi:10.1016/j.physletb.2007.11.065
  [arXiv:0710.0325 [hep-ph]].

\bibitem{Matsuki:2004zu}
  T.~Matsuki, K.~Mawatari, T.~Morii and K.~Sudoh,
  hep-ph/0408326.

\bibitem{Matsuki:2007fb}
  T.~Matsuki and K.~Seo,
  Prog.\ Theor.\ Phys.\  {\bf 118}, 1087 (2007)
  [Prog.\ Theor.\ Phys.\  {\bf 121}, 1141 (2009)]
  doi:10.1143/PTP.121.1141, 10.1143/PTP.118.1087
  [hep-ph/0703158 [HEP-PH]].


\end{thebibliography}
\end{document}